\documentclass[a4paper,11pt]{article}
\usepackage{pos}

\DeclareUnicodeCharacter{03BC}{\ensuremath{\mu}}
\DeclareUnicodeCharacter{03B7}{\ensuremath{\eta}}

\usepackage{subcaption}
\usepackage{booktabs}
\usepackage{listings}
\usepackage{listingsutf8}
\usepackage{fvextra}
\usepackage{multicol}
\usepackage{tabularx}
\usepackage{cleveref}
\usepackage{caption}

\renewcommand{\logo}{\relax}

\crefname{figure}{Figure}{Figures}
\Crefname{figure}{Figure}{Figures}
\crefname{table}{Table}{Tables}
\Crefname{table}{Table}{Tables}
\crefname{section}{Section}{Sections}
\Crefname{section}{Section}{Sections}
\crefname{listing}{Listing}{Listings}
\Crefname{listing}{Listing}{Listings}

\title{Development of an LLM-Based System for Automatic Code Generation from HEP Publications}
\ShortTitle{Automatic Code Generation from HEP Publications}

\author*[a]{Masahiko Saito}
\author[b]{Tomoe Kishimoto}
\author[a]{Junichi Tanaka}

\affiliation[a]{International Center for Elementary Particle Physics (ICEPP), The University of Tokyo,\\
  7-3-1, Hongo, Bunkyo, Tokyo, Japan}

\affiliation[b]{High Energy Accelerator Research Organization (KEK), Computing Research Center,\\
  1-1, Oho, Tsukuba, Ibaraki, Japan}

\emailAdd{saito@icepp.s.u-tokyo.ac.jp}

\abstract{
Ensuring the reproducibility of physics results is one of the crucial challenges in high-energy physics~(HEP).
In this study, we develop a proof-of-concept system that uses large language models~(LLMs) to extract analysis procedures from HEP publications and generate executable analysis code for reproducing published results. 

Our method consists of two stages.
In the first stage, open-weight LLMs extract event selection criteria, object definitions, and other relevant analysis information from a target paper and, when necessary, from its referenced publications, and then produce a structured selection list.
In the second stage, the structured selection list is used to generate analysis code, which is then executed and validated iteratively. 

As a benchmark, we use the ATLAS $H \rightarrow ZZ^{*} \rightarrow 4\ell$ analysis based on proton-proton collision data recorded in 2015 and 2016 and released as ATLAS Open Data.
This benchmark allows direct comparison between the generated results and the published analysis, as well as comparison with a manually developed baseline implementation.
We separately evaluate selection extraction and code generation in order to clarify the current capabilities and limitations of open-weight LLMs for HEP analysis reproduction.

Our initial results show that recent open-weight models can recover many documented selection criteria from papers and references, and that in some runs they can generate event selections fully matching a baseline implementation at the event level.
At the same time, stochasticity, hallucination, and execution failure remain significant challenges.
These results suggest that LLMs are already promising as human-in-the-loop tools for reproducibility support, although they are not yet reliable as fully autonomous HEP analysis agents.
In this paper, we report the design of the prototype system and its initial performance evaluation. 

}

\FullConference{International Symposium on Grids and Clouds (ISGC2026)\\
15-20, March, 2026 \\
Academia Sinica, Taipei, Taiwan\\}

\begin{document}
\maketitle

\section{Introduction}
Data analysis in high energy physics~(HEP) has become increasingly complex, demanding substantial computational expertise and time to set up environments and write code. This situation raises the barrier to entry for students and newcomers.

Large language models~(LLMs) offer powerful support for coding and analysis frameworks, potentially allowing researchers to focus more on underlying physics problems.
However, due to stochastic variation and hallucination, fully automated analyses remain difficult to trust without careful verification.
Therefore, a human-in-the-loop framework is essential for applying LLMs to analysis support.

\begin{figure}[htbp]
  \begin{center}
    \includegraphics[width=0.6\columnwidth]{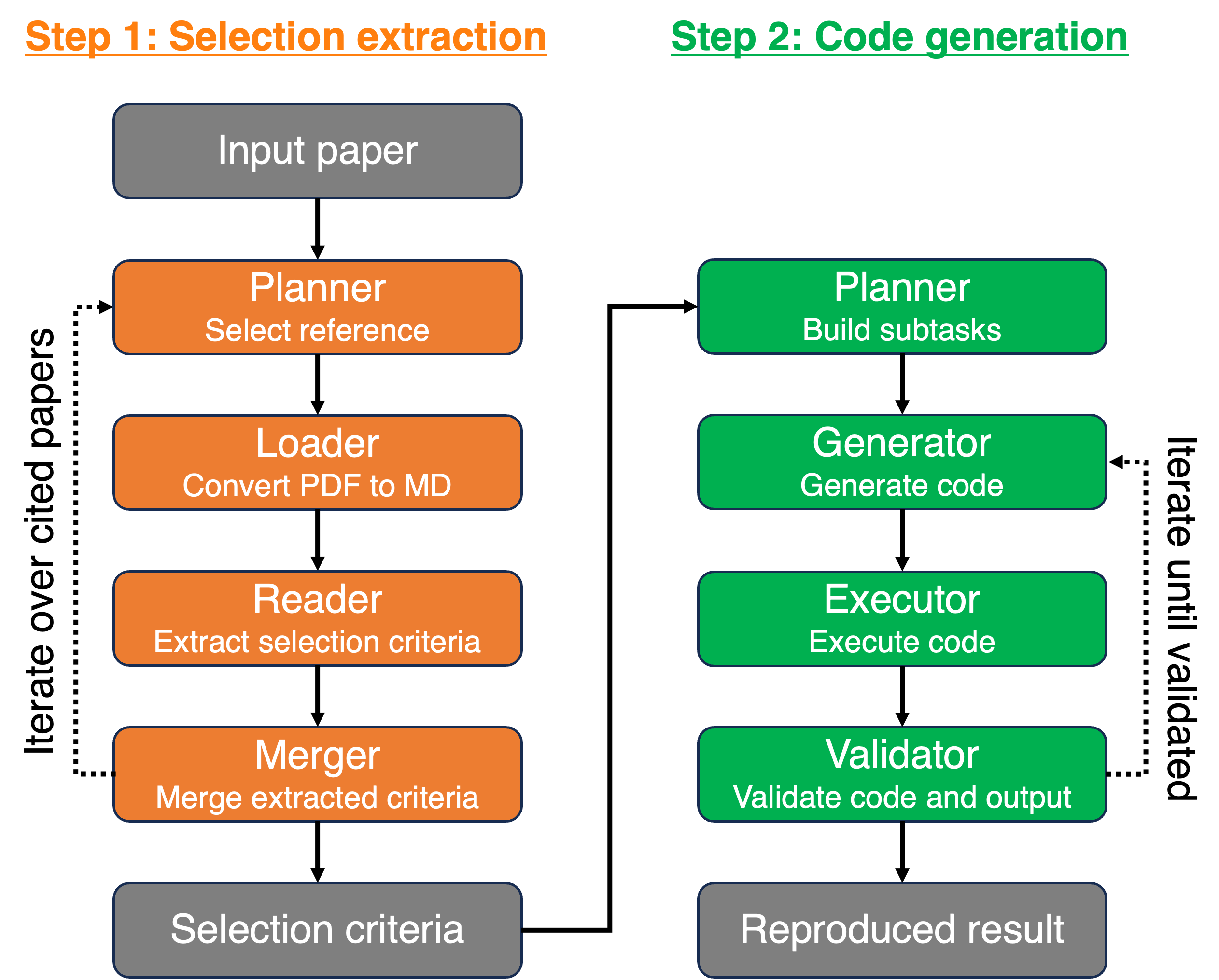}
    \caption{Overview of the proposed two-stage workflow. Step 1 iteratively extracts and merges selection criteria from the target paper and its references. Step 2 utilizes these extracted criteria to sequentially generate, execute, and validate analysis code until successful reproduction is achieved.}
    \label{fig:overview}
  \end{center}
\end{figure}

In this study, we develop a proof-of-concept~(PoC) LLM workflow pipeline that extracts analysis procedures from HEP publications and automatically generates executable analysis code~(\cref{fig:overview}).
The proposed method consists of two main stages implemented with open-weight LLMs.
First, the system extracts selection criteria from the paper PDF and cited references to iteratively organize a structured selection list.
Second, this list is used as input to iteratively generate, execute, and verify analysis code.

Rather than directly generating code from the text, our approach introduces a human-readable intermediate representation, positioning the LLM as a verifiable collaborator rather than a black-box system.
Furthermore, this system functions as a framework for assessing the reproducibility of HEP publications; successful reproduction suggests sufficient documentation, while failure may indicate missing or ambiguous descriptions.
In the longer term, this tool could be used before publication to improve paper quality.

Our main contributions are twofold.
First, we implement a workflow that extracts structured selection criteria from papers and iteratively generates, executes, and validates analysis code.
Second, using the ATLAS Open Data $H \rightarrow ZZ^{*} \rightarrow 4\ell$ analysis as a benchmark, we separately evaluate paper understanding and code generation to quantitatively clarify the strengths and limitations of open-weight LLMs.

The remainder of this paper details related work (\cref{sec:related_work}), the benchmark design (\cref{sec:benchmark}), selection extraction (\cref{sec:step1}), code generation (\cref{sec:step2}), and conclusions with future directions (\cref{sec:discussion}, \cref{sec:conclusion}).

\section{Related work} \label{sec:related_work}

In recent years, large language models~(LLMs) have been increasingly applied to HEP workflow automation. 
For instance, Esmail et al.~\cite{2602.06496} developed CoLLM, an end-to-end analysis environment using open-weight models. 
Gendreau-Distler et al.~\cite{2512.07785} proposed an agent-based framework integrated with Snakemake for ATLAS Open Data analysis, while Moreno et al.~\cite{2603.20179} demonstrated end-to-end execution of multiple analyses using Claude Code and literature-based retrieval.

Unlike these studies, which focus on general end-to-end automation, our work specifically aims to extract analysis procedures directly from existing HEP publications to quantitatively evaluate reproducibility. 
Furthermore, we distinguish our approach by explicitly separating the workflow into verifiable document understanding and code generation stages using LangChain and LangGraph, while focusing primarily on the capabilities of open-weight models.

\section{Benchmark Design and Evaluation Protocol} \label{sec:benchmark}

As our benchmark, we selected the ATLAS $H \to ZZ^* \to 4\ell$ analysis~\cite{2018345}.
Because this paper delegates detailed event selection definitions to cited references, it is ideal for testing an LLM's ability to trace and integrate multi-document information.
The evaluation utilizes 2015--2016 DAOD\_PHYSLITE ATLAS Open Data~\cite{atlas_opendata_data_2015,atlas_opendata_data_2016,atlas_opendata_mc_higgs,atlas_opendata_mc_ewboson}.

\begin{figure}[htbp]
  \centering
  \begin{subfigure}[t]{0.48\columnwidth}
    \centering
    \includegraphics[width=\linewidth, trim=0 30 0 0, clip]{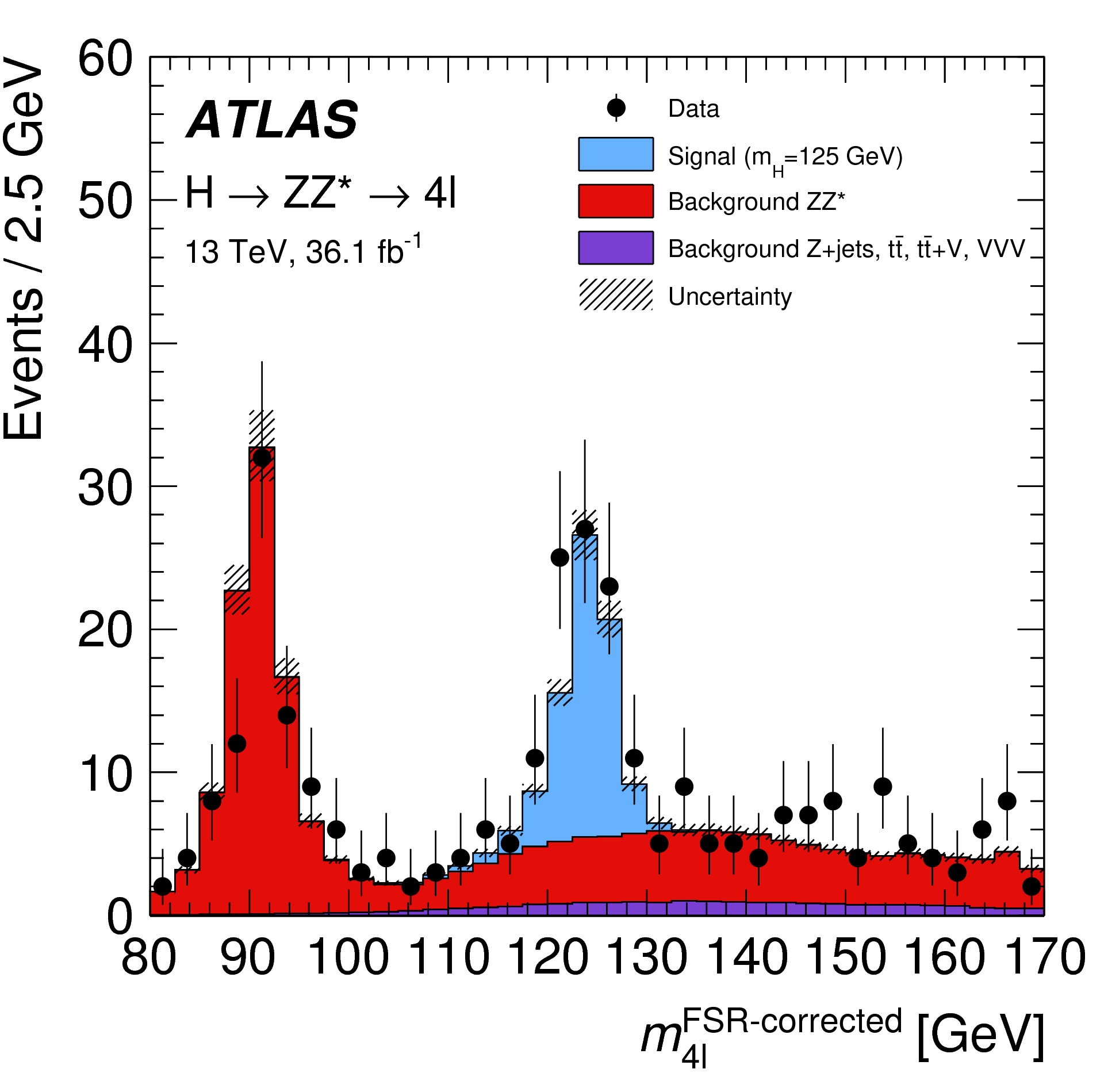}
    \caption{Published paper~\cite{170802810}}
  \end{subfigure}
  \begin{subfigure}[t]{0.47\columnwidth}
    \centering
    \includegraphics[width=\linewidth]{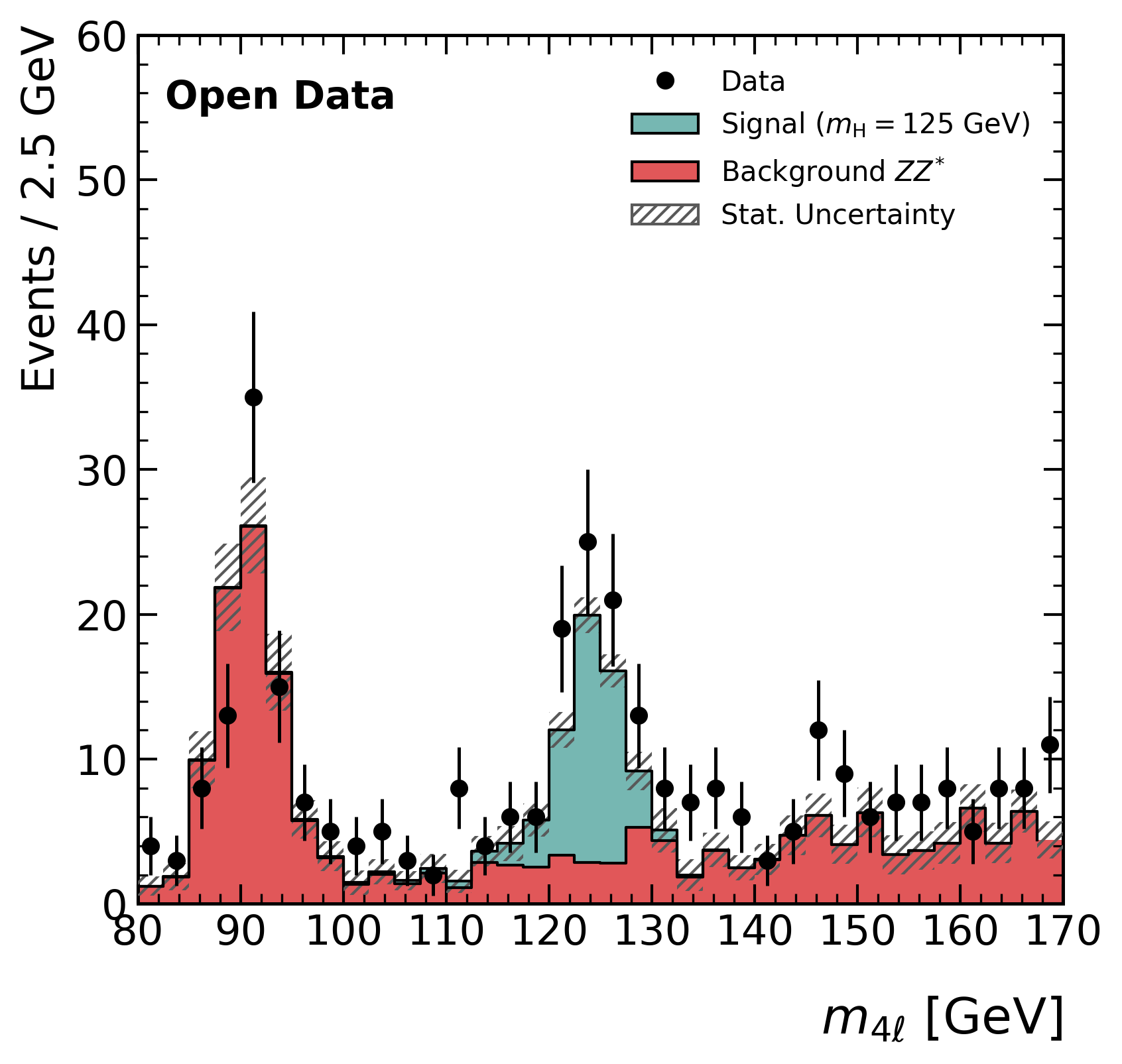}
    \caption{Manually reproduced distribution}
  \end{subfigure}
  \caption{Comparison of the four-lepton invariant mass distribution between (a) the published result~\cite{170802810} and (b) our manually reproduced baseline.}
  \label{fig:fourleptonmass}

\end{figure}

We manually reproduced the analysis to establish a baseline.
While missing variables in the Open Data and omitted paper details prevented a perfect match with the published distribution~(\cref{fig:fourleptonmass}), the overall result was reasonably consistent.
We use this manual implementation and a curated list of 27 explicitly identifiable selection criteria as our ground truth~(Appendix~\ref{appendix:benchmark}).

To clearly identify performance bottlenecks, we evaluate the two workflow stages separately.
Step~1 (selection extraction) is evaluated against the 27 ground-truth cuts, using the number of correctly extracted cuts and contradictory hallucinations as metrics.
Step~2 (code generation) evaluates only the cuts implementable via the Open Data Ntuple.
The generated code is compared to the baseline implementation at the event level, categorizing results as \textit{Exactly Matched} (identical event list), \textit{Not Matched} (valid execution but differing events), or \textit{Execution Failed}.

\section{Step 1: Selection Extraction} \label{sec:step1}

The goal of Step~1 is to extract event-selection criteria and object definitions from the target paper and its cited references.
Because critical definitions are often delegated to citations, this multi-document extraction is essential for completeness. 
Rather than a simple numerical list, the extracted information is maintained as a structured representation containing comments and reference provenance~(Appendix~\ref{appendix:structured_output}).
This format facilitates downstream information integration and provides a verifiable, human-readable intermediate result.

\subsection{Workflow Design}

The selection extraction system is an iterative workflow consisting of four components~(\cref{fig:overview}). 
First, the \textit{Planner} determines the next reference to read and formulates a specific reading objective based on the current selection list.
This targeted approach prevents the LLM from extracting noisy or irrelevant information from supplementary references. 
Next, the \textit{Loader} converts the PDF to Markdown using \texttt{marker}~\cite{marker}, uses an LLM to isolate relevant body text, and maps in-text citations to arXiv IDs.
To ensure practical runtimes, reference tracing is restricted to directly cited ATLAS papers available on arXiv.
The \textit{Reader} then extracts selection criteria based on the Planner's objective.
We test a \textit{Bulk} mode (processing the full text at once) and a \textit{Chunk} mode (processing sequential segments for models with limited context windows).
To mitigate context-loss errors in Chunk mode, an LLM filter is applied to each chunk, and only task-relevant chunks are passed to the Reader. 
Finally, the \textit{Merger} integrates the new results.
To prevent inappropriately overwriting primary information, it treats cited references as supplementary and updates the existing selection list only when current descriptions are ambiguous.

\subsection{Experimental Setup}

\begin{table}[t]
\centering
\small
\begin{tabular}{lccc|c}
\toprule
Model & Params (B) & Float length (bit) & Release date & Target Step \\
\midrule
Qwen3:4B   & 4   & 16 & 2025/09/17 & Step 1 \\
Qwen3:30B  & 31  & 16 & 2025/09/17 & Step 1 \\
Qwen3:235B & 235 & 8  & 2025/09/17 & Step 1 \\
Qwen3-Coder:30B  & 31  & 16 & 2025/12/03 & Step 2 \\
Qwen3-Coder-Next  & 80  & 8 & 2026/02/03 & Step 2 \\
GPT-OSS:120B & 120 & 4 & 2025/08/27 & Step 1, 2 \\
Gemini 2.5 Flash-Lite & - & - & 2025/07/22 & Step 1 \\
Gemini 2.5 Flash & - & - & 2025/06/17 & Step 1 \\
\bottomrule
\end{tabular}
\caption{Summary of the LLMs evaluated in this study, including model size, numerical precision, release date, and the step in which they were used.}
\label{tab:llm_summary}
\end{table}

The extraction workflow is implemented using LangChain~\cite{Chase_LangChain_2022}, LangGraph~\cite{LangGraph}, and vLLM~\cite{kwon2023efficient} (code will be made publicly available at \url{https://github.com/saitoicepp/isgc2026_ai_workflow}). 
We primarily evaluate open-weight models (Qwen3~\cite{qwen3} and GPT-OSS~\cite{openai2025gptoss120bgptoss20bmodel}) due to their practical advantages, such as data privacy and cost-efficiency.
Commercial models (Gemini 2.5 Flash-Lite and Flash~\cite{comanici2025gemini}) are included as baselines~(\cref{tab:llm_summary}).

To assess stochasticity, we executed each model 10 times (max 20K completion tokens), treating invalid structured outputs as failures.
Computations used a single Nvidia A100 (80~GB), except for Qwen3:235B, which required eight A100s (40~GB).

For consistent automated evaluation, an LLM-based judge compared the generated cut lists against the ground truth.
To minimize variance, the final metrics (correct cuts and hallucinations) are calculated as the median of six evaluations (three runs each by Qwen3:30B and GPT-OSS:120B judges).

\subsection{Results}

\begin{figure}[htbp]
  \centering
  \begin{subfigure}{\columnwidth}
    \centering
    \includegraphics[width=0.85\linewidth]{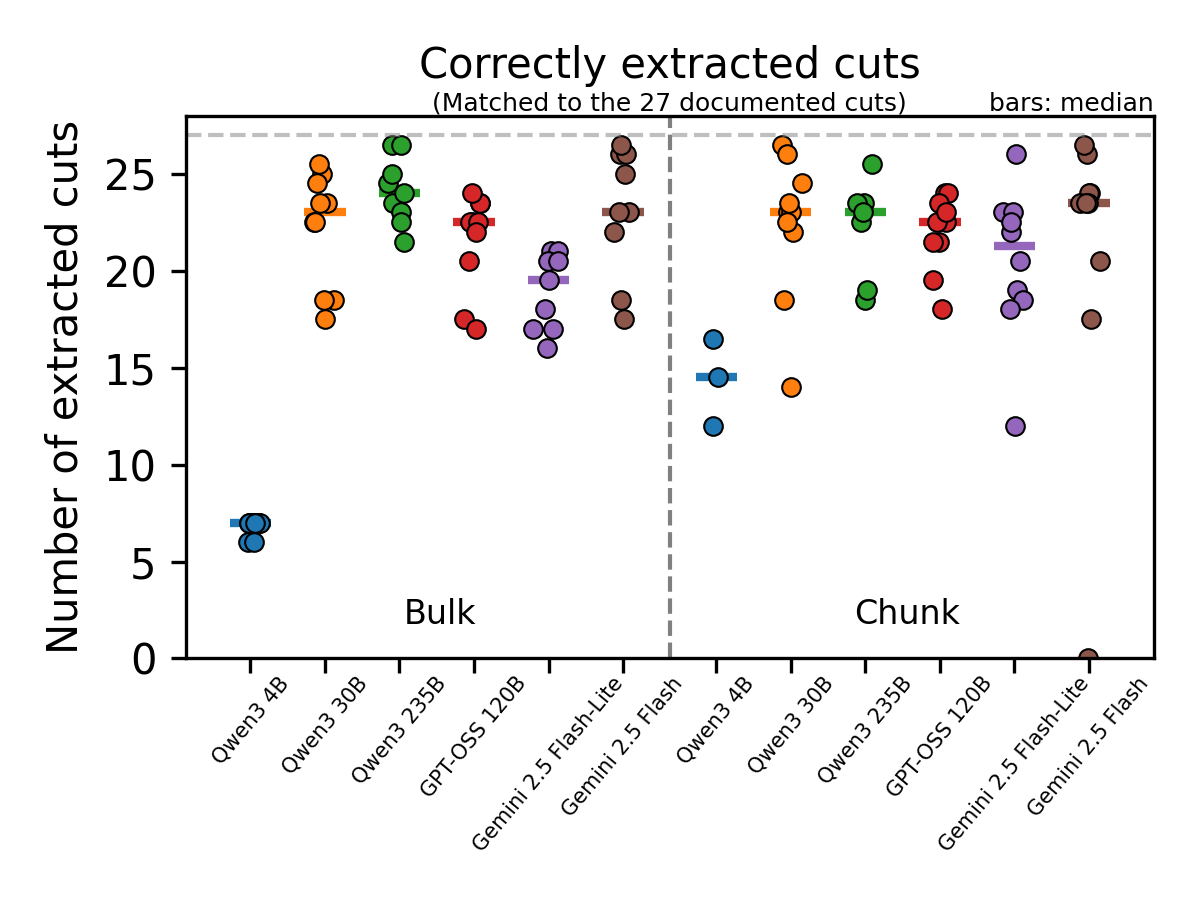}
    \caption{Correctly extracted cuts}
  \end{subfigure}
  \begin{subfigure}{\columnwidth}
    \centering
    \includegraphics[width=0.85\linewidth]{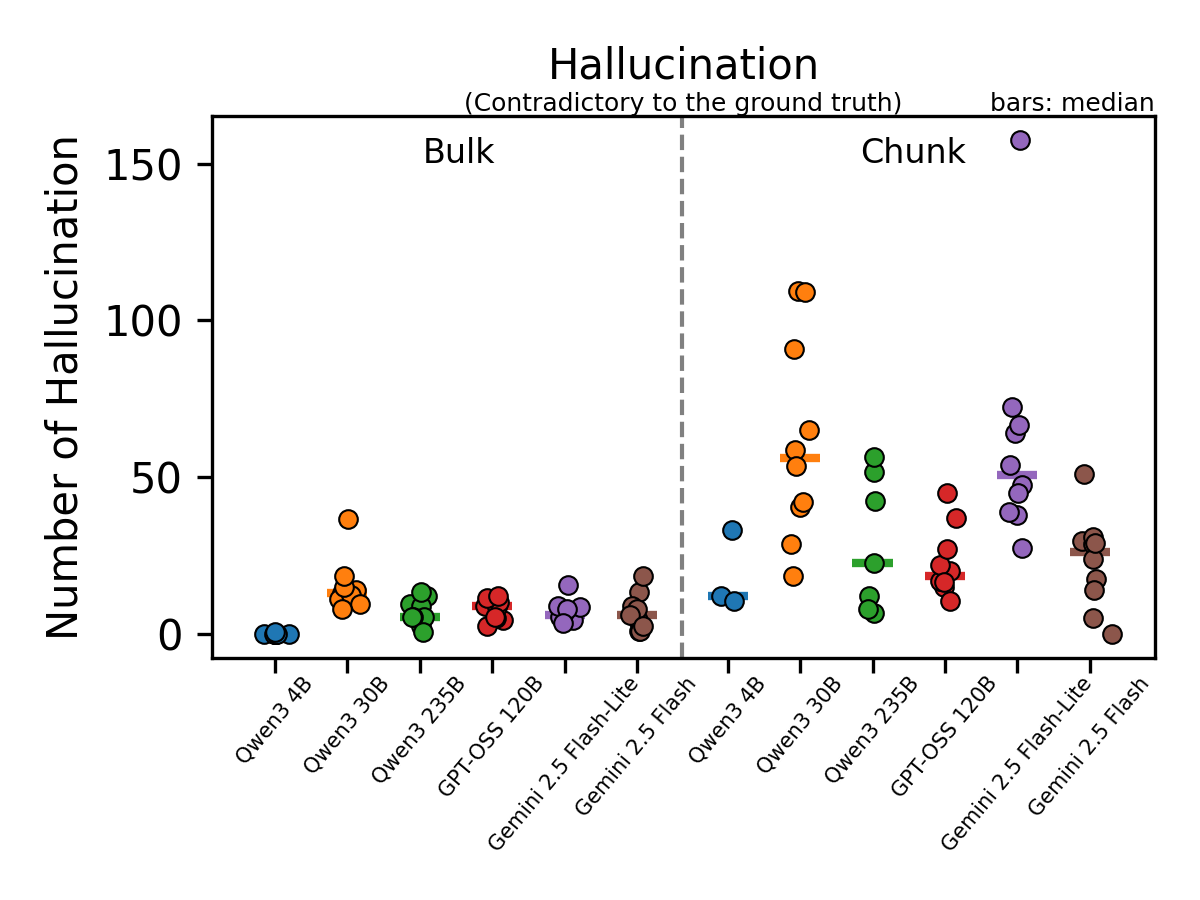}
    \caption{Hallucinations}
  \end{subfigure}
  \caption{Comparison of Bulk and Chunk settings in Step 1: (a) correctly extracted cuts (out of 27 ground-truth cuts) and (b) hallucinations. Points represent successful runs, with horizontal bars indicating medians. Failed runs are omitted.}
  \label{fig:step1_result}
\end{figure}

\Cref{fig:step1_result} shows the selection extraction results, comparing the Bulk (processing the full text at once) and Chunk (processing sequential segments) settings.

In the Bulk setting (left side of each panel), models with $\ge$30B parameters extracted most documented cuts, with Qwen3:235B and Gemini 2.5 Flash successfully identifying all 27 cuts in some runs.
Conversely, the 4B model performed poorly.
However, all models exhibited substantial run-to-run variation, and while larger models produced fewer hallucinations, contradictory statements were never entirely eliminated.
This persistent stochasticity highlights the necessity of human verification.

Regarding the Chunk setting (right side of each panel), sequentially processing text segments for Qwen3:4B recovered more correct cuts than the Bulk setting, suggesting localized context aids retrieval for models with limited capacity.
However, this came at a severe cost: the Chunk setting drastically increased both hallucinations and the workflow failure rate (from $0/10$ to $7/10$).
This degradation likely occurs because more LLM calls create more opportunities for invalid outputs.

\section{Step 2: Code Generation} \label{sec:step2}

The goal of Step~2 is to generate analysis code from the structured selection criteria obtained in Step~1.
Specifically, the task is to generate code that matches the baseline event selection for ATLAS $H \to ZZ^{*} \to 4\ell$ MC events, evaluating agreement at the event level.

Currently, the system lacks autonomous mechanisms, such as RAG, to retrieve HEP-specific domain knowledge like API usage and variable definitions.
Therefore, we evaluate code generation under a controlled setting where the prompt explicitly provides all necessary specifications, including available variables and execution constraints.
Consequently, Step~2 is not an evaluation of fully autonomous end-to-end analysis, but an isolated assessment of the LLM's ability to translate structured criteria into executable code, leaving autonomous knowledge retrieval for future work.

\subsection{Workflow Design}

The code generation system is an iterative LangGraph-based workflow comprising four components~(\cref{fig:overview}). To support the LLM, the prompt explicitly provides the task description, structured selection criteria, available Ntuple variables, and execution constraints.

The \textit{Planner} decomposes the overall task into sequential subtasks and defines concrete completion criteria for each. 
The \textit{Generator} produces code for the current subtask. To iteratively refine the implementation and avoid repeating mistakes, it receives the completion criteria, previous validation feedback, runtime errors, and the most recent successful and failed code snippets. 
The \textit{Executor} securely runs the generated code within an isolated Singularity container pre-configured with tools such as ROOT, numpy, and uproot, passing execution logs to the Validator. 
Finally, the \textit{Validator} evaluates both the execution results and the code itself against the completion criteria. This dual check ensures the code genuinely implements the intended procedure rather than merely suppressing runtime errors.
If validation fails, the Generator produces revised code; if successful, the system advances to the next subtask.

\subsection{Experimental Setup}

As in Step~1, the workflow utilizes LangChain, LangGraph, and vLLM. We evaluated LLMs specialized for code generation~(\cref{tab:llm_summary}), executing each 10 times (max 20K completion tokens) to assess stochasticity.
Qwen3-Coder:30B and GPT-OSS:120B were run on a single Nvidia A100 (80~GB), while Qwen3-Coder-Next used two.

To ensure practical computation times, the evaluation was limited to 1,000 MC $H \to ZZ^* \to 4\ell$ events.
We evaluated event-level agreement by comparing the output of the generated code against the manual baseline, which selected 235 out of the 1,000 events.
Following the criteria defined in \cref{sec:benchmark}, results are categorized as \textit{Exactly Matched} (perfect agreement with the 235 baseline events), \textit{Not Matched}, or \textit{Execution Failed} (subtask unresolved after 10 attempts).

\subsection{Results}

\Cref{tab:step2_result} summarizes the 10-run performance.
Qwen3-Coder-Next:80B and GPT-OSS:120B achieved exact event-level agreement with the baseline in 3/10 and 2/10 runs, respectively, whereas Qwen3-Coder:30B produced zero exact matches.
This demonstrates that while open-weight LLMs can successfully generate baseline-matching HEP code, the outcome is highly dependent on the specific model used.

However, the high frequency of \textit{Not Matched} and \textit{Execution Failed} outcomes across all models indicates that current stability and correctness remain insufficient.
Crucially, the existence of executable code that yields incorrect event selections highlights that execution success is not a proxy for physical correctness, emphasizing the strict need for event-level verification.
While this structured generation approach is promising, improving reliability is a key challenge for future work.

\begin{table}[h]
\centering
\begin{tabular}{lccc}
\toprule
Model & Exactly Matched & Not Matched & Execution Failed \\
\midrule
Qwen3-Coder-Next:80B & 3 & 2 & 5 \\
Qwen3-Coder:30B      & 0 & 5 & 5 \\
GPT-OSS:120B         & 2 & 1 & 7 \\
\bottomrule
\end{tabular}
\caption{Step 2 performance over 10 runs, categorized by event-level agreement with the reference implementation.}
\label{tab:step2_result}
\end{table}

\section{Discussions and Future Directions} \label{sec:discussion}

This study highlights several fundamental challenges in LLM-assisted HEP reproduction.
First, PDF-to-text conversion remains unstable for complex layouts, creating a preprocessing bottleneck independent of LLM capabilities.
Second, persistent stochasticity and hallucinations across both stages dictate that superficially plausible outputs cannot be blindly trusted.
Rigorous verification mechanisms, such as structured reference tracking and sandboxed execution, are essential.
Consequently, a human-in-the-loop workflow, where physicists inspect verifiable intermediate states, remains the most practical approach, positioning the LLM as a collaborative tool rather than a fully autonomous analyst.

Future work will prioritize an end-to-end evaluation to quantify how selection extraction errors in Step~1 propagate to code generation in Step~2.
Furthermore, integrating RAG for HEP-specific domain knowledge (e.g., ROOT and Awkward Array APIs) should help mitigate execution failures and selection mismatches.
We also plan to expand the benchmark beyond the $H \to ZZ^* \to 4\ell$ analysis to assess broader generalization.
Ultimately, we aim to develop a collaborative framework that not only translates text to code but also explicitly identifies ambiguities in published procedures, thereby helping to improve the overall reproducibility and transparency of HEP literature.

\section{Conclusion} \label{sec:conclusion}

In this study, we developed a verifiable, LLM-based proof-of-concept system that extracts structured analysis procedures from HEP publications and generates executable code.
Using the ATLAS $H\to ZZ^* \to 4\ell$ Open Data analysis as a benchmark, we demonstrated that while open-weight models with $\ge$30B parameters can successfully extract complex criteria and generate baseline-matching code, persistent stochasticity, hallucinations, and execution failures remain significant challenges. 

Consequently, current open-weight LLMs are not yet reliable enough to act as fully autonomous analysis agents, but they show strong promise as human-in-the-loop collaborative tools.
Future work will focus on end-to-end evaluation, expanded benchmarks, and RAG integration to build a more robust, practical analysis-support system.

\acknowledgments
We would like to thank Yusuke Oda for valuable advice and helpful discussions on this work.
This work is partially funded by grants 22H05113, ``Foundation of Machine Learning Physics.''
We acknowledge the work of the ATLAS Collaboration to record or simulate, reconstruct, and distribute the Open Data used in this paper, and to develop and support the software with which it was analysed.

\begingroup
\raggedright
\bibliographystyle{JHEPcustom}
\bibliography{main}
\endgroup

\appendix

\section{Benchmark dataset}\label{appendix:benchmark}

\Cref{truth_selection_list} summarizes the event selection criteria used as the ground truth for Step~1 extraction and the input specification for Step~2 code generation.
We compiled these cuts (pre-selection, electron, muon, and quadruplet) based on explicit descriptions in the main text and cited references.
Entries marked with (*) are excluded from both evaluations, as they cannot be accurately reproduced due to ambiguous definitions in the literature or missing variables in the ATLAS Open Data Ntuple.

\begin{table}[htbp]
\centering
\small
\begin{tabular}{ll}
\toprule
Step & Description \\
\midrule
\multicolumn{2}{l}{\textbf{1. Pre-selection}} \\
1.1 & (*) Good Run List \\
1.2 & (*) Trigger requirements \\
1.3 & Number of primary vertices $> 0$ \\

\multicolumn{2}{l}{\textbf{2. Electron}} \\
2.1 & Loose criteria \\
2.2 & $E_{\mathrm{T}} > 7$ GeV \\
2.3 & $|\eta| < 2.47$ \\
2.4 & $(p_{\mathrm{T}}^{\mathrm{cone20}} / E_{\mathrm{T}}) < 0.15$ \\
2.5 & $(E_{\mathrm{T}}^{\mathrm{cone20}} / E_{\mathrm{T}}) < 0.20$ \\
2.6 & $|z_0 \sin\theta |< 0.5$ mm \\
2.7 & $|d_0 / \sigma(d_0)| < 5$ \\

\multicolumn{2}{l}{\textbf{3. Muon}} \\
3.1 & $|\eta| < 2.7$ \\
3.1.1 & $|\eta| < 0.1$ for Segmented-tagged muon and Calo-tagged muon \\
3.1.2 & $0.1 < |\eta| < 2.5$ for Combined muons \\
3.1.3 & $2.4 < |\eta| < 2.7$ for Muon-Spectrometer standalone muons\\
3.2 & $p_{\mathrm{T}} > 5$ GeV \\
3.2.1 & $p_{\mathrm{T}} > 15~\mathrm{GeV}$ for Calo-tagged muon \\
3.3 & $p_{\mathrm{T}}^{\mathrm{cone30}}/p_{\mathrm{T}} < 0.15$ \\
3.4 & $E_{\mathrm{T}}^{\mathrm{cone20}}/p_{\mathrm{T}} < 0.30$ \\
3.5 & $|z_0 \sin\theta |< 0.5~\mathrm{mm}$ \\
3.6 & $|d_0/\sigma(d_0)| < 3$ \\
3.7 & $|d_0| < 1~\mathrm{mm}$ \\

\multicolumn{2}{l}{\textbf{4. Quadruplet}} \\
4.1 & Number of same-flavour opposite-sign lepton pairs $\geq 2$ \\
4.2 & $50 < m_{12} < 106~\mathrm{GeV}$ \\
4.3 & $12 < m_{34} < 115~\mathrm{GeV}$ \\
4.4 & $p_{\mathrm{T}} > 20,\ 15,\ 10~\mathrm{GeV}$ for 1st, 2nd, 3rd leptons \\
4.5 & $\Delta R(\ell,\ell) > 0.1\ (0.2)\ \text{for same-flavour (opposite-flavour)}$ \\
4.6 & $m_{\ell\ell} > 5~\mathrm{GeV}$ for same-flavour opposite-sign leptons \\
4.7 & (*) 4 leptons vertex fit \\
4.8 & Number of Combined muons $\geq 3$ for $4\mu$ channel \\
4.9 & (*) $Z$ mass constraint \\
4.10 & (*) FSR correction \\
4.11 & $110 < m_{4\ell} < 135~\mathrm{GeV}$ \\

\bottomrule
\end{tabular}
\caption{Event selection criteria used in the benchmark. Asterisks (*) denote cuts difficult to reproduce due to ambiguous definitions or missing Open Data variables.}
\label{truth_selection_list}
\end{table}

\section{Example of structured object used in a workflow}\label{appendix:structured_output}

\Cref{lst:step1_markdown_example} illustrates the input Markdown text converted from the PDF via \texttt{marker}, preserving essential document structure and in-text citations.

\begin{lstlisting}[
  caption={Input Markdown text converted from the target paper~\cite{2018345} (Sec.~7.1).},
  label={lst:step1_markdown_example},
  breaklines=true,
  columns=fullflexible,
  breakatwhitespace=true,
  frame=single,
  % basicstyle=\ttfamily\small,
  basicstyle=\ttfamily\tiny,
]
# 7 Mass measurement in the $H \rightarrow ZZ^* \rightarrow 4\ell$ channel
### 7.1 Event selection
Events are required to contain at least four isolated leptons ( $\ell = e, \mu$ ) that emerge from a common vertex, form two pairs of oppositely charged same-flavour leptons. Electrons are required to be within the full pseudorapidity range of the inner tracking detector ( $|\eta| < 2.47$ ) and have transverse energy $E_T > 7$ GeV, while muons are required to be within the pseudorapidity range of the muon spectrometer ( $|\eta| < 2.7$ ) and have transverse momentum $p_T > 5$ GeV. The three higher- $p_T$ ( $E_T$ ) leptons in each quadruplet are required to pass thresholds of 20, 15, and 10 GeV, respectively. A detailed description of the event selection can be found in Ref. [11, 44].
\end{lstlisting}

From this text, the Reader extracts a structured JSON object organized by category; an example is shown in \cref{lst:structured_object_example}.

Each entry captures the cut itself (\texttt{content}), supplementary context (\texttt{comments}), and references to trace for further details (\texttt{references}).
To enable downstream literature tracing by the Planner and Loader, standard in-text citation IDs (e.g., [11, 44]) are explicitly converted to arXiv IDs within the \texttt{references} field.
This format ensures the extracted criteria remain fully verifiable for both the automated workflow and human inspection.

\begin{center}
\captionsetup{type=lstlisting}
\captionof{lstlisting}{Example of the structured JSON object generated by the Reader.}
\label{lst:structured_object_example}
\begin{Verbatim}[
  breaklines=true,
  breakanywhere=true,
  fontsize=\scriptsize,
  % fontsize=\tiny,
  frame=single
]
{
  "selection": {
    "Event": [
      {
        "content": "At least four isolated leptons ($\ell$ = e, μ) from a common vertex",
        "comments": "Leptons must form two pairs of oppositely charged same-flavour leptons.",
        "references": [
          "1708.02810",
          "1712.02304"
        ]
      },
      {
        "content": "Three higher-pT (E_T) leptons in each quadruplet pass thresholds of 20, 15, and 10 GeV, respectively",
        "comments": "Applies to the three highest transverse energy/momentum leptons in the event.",
        "references": [
          "1708.02810",
          "1712.02304"
        ]
      }
    ],
    "Electron": [
      {
        "content": "|η| < 2.47",
        "comments": "Full pseudorapidity range of the inner tracking detector.",
        "references": [
          "1708.02810",
          "1712.02304"
        ]
      },
      {
        "content": "E_T > 7 GeV",
        "comments": "Transverse energy threshold for electron selection.",
        "references": [
          "1708.02810",
          "1712.02304"
        ]
      }
    ],
    "Muon": [
      {
        "content": "|η| < 2.7",
        "comments": "Pseudorapidity range of the muon spectrometer.",
        "references": [
          "1708.02810",
          "1712.02304"
        ]
      },
      {
        "content": "p_T > 5 GeV",
        "comments": "Transverse momentum threshold for muon selection.",
        "references": [
          "1708.02810",
          "1712.02304"
        ]
      }
    ]
  }
}
\end{Verbatim}
\end{center}

\end{document}